\documentclass[twocolumn, prx,reprint, amsmath,amssymb,showpacs,superscriptaddress,longbibliography]{revtex4-1}

\usepackage{graphicx}
\usepackage{dcolumn}
\usepackage{bm}
\usepackage{graphicx}
\usepackage{epsfig}
\usepackage{epsf}
\usepackage{amssymb}
\usepackage{amsmath}
\usepackage{amsthm}
\usepackage{multirow}
\usepackage{cases}
\usepackage[colorlinks=true,linkcolor=blue,citecolor=blue,pdfauthor={ },pdftitle={ },pdfsubject={ },pdfkeywords={ }]{hyperref}
\usepackage{pgfplots}
\usepackage{lipsum}

\begin{document}

\title{Measuring out-of-time-order correlators on a nuclear magnetic resonance quantum simulator}

\author{Jun Li}
\affiliation{Beijing Computational Science Research Center, Beijing 100193, China}

\author{Ruihua Fan}
\affiliation{Institute for Advanced Study, Tsinghua University, Beijing, 100084, China}
\affiliation{Department of Physics, Peking University, Beijing, 100871, China}

\author{Hengyan Wang}
\affiliation{Hefei National Laboratory for Physical Sciences at Microscale and Department of Modern Physics, University of Science and Technology of China, Hefei, Anhui 230026, China}

\author{Bingtian Ye}
\affiliation{Department of Physics, Peking University, Beijing, 100871, China}

\author{Bei Zeng}
\email{zengb@uoguelph.ca}
\affiliation{Department of Mathematics \& Statistics, University of Guelph, Guelph N1G 2W1, Ontario, Canada}
\affiliation{Institute for Quantum Computing, University of Waterloo, Waterloo N2L 3G1, Ontario, Canada}
\affiliation{Institute for Advanced Study, Tsinghua University, Beijing, 100084, China}

\author{Hui Zhai}
\email{zhai@tsinghua.edu.cn}
\affiliation{Institute for Advanced Study, Tsinghua University, Beijing, 100084, China}
\affiliation{Collaborative Innovation Center of Quantum Matter, Beijing, 100084, China}

\author{Xinhua Peng}
\email{xhpeng@ustc.edu.cn}
\affiliation{Hefei National Laboratory for Physical Sciences at Microscale and Department of Modern Physics, University of Science and Technology of China, Hefei, Anhui 230026, China}
\affiliation{Synergetic Innovation Centre of Quantum Information $\&$ Quantum Physics, University of Science and Technology of China, Hefei, Anhui 230026, China}
\affiliation{College of Physics and Electronic Science, Hubei Normal University, Huangshi, Hubei 435002, China}

\author{Jiangfeng Du}
\affiliation{Hefei National Laboratory for Physical Sciences at Microscale and Department of Modern Physics, University of Science and Technology of China, Hefei, Anhui 230026, China}
\affiliation{Synergetic Innovation Centre of Quantum Information $\&$ Quantum Physics, University of Science and Technology of China, Hefei, Anhui 230026, China}

\begin{abstract}
The idea of the out-of-time-order correlator (OTOC) has recently emerged in the study of both condensed matter systems and gravitational systems. It not only plays a key role in investigating the holographic duality between a strongly interacting quantum system and a gravitational system, but also diagnoses the chaotic behavior of many-body quantum systems and characterizes the information scrambling.
Based on the OTOCs, three different concepts -- quantum chaos, holographic duality, and information scrambling -- are found to be intimately related to each other. Despite of its theoretical importance, the experimental measurement of the OTOC is quite challenging and so far there is no experimental measurement of the OTOC for local operators. Here we report the measurement of OTOCs of local operators for an Ising spin chain on a nuclear magnetic resonance quantum simulator. We observe that the OTOC behaves differently in the integrable and non-integrable cases.
Based on the recent discovered relationship between OTOCs and the growth of entanglement entropy in the many-body system, 
we extract the entanglement entropy from the measured OTOCs, which clearly shows that the information entropy oscillates in time for integrable models and scrambles for non-intgrable models. 
With the measured OTOCs, we also obtain the experimental result of the butterfly velocity, which measures the speed of correlation propagation. 
Our experiment paves a way for experimentally studying quantum chaos, holographic duality, and information scrambling in many-body quantum systems with quantum simulators.
\end{abstract}


\maketitle

\section{Introduction}

The out-of-time-order correlator (OTOC), given by
\begin{equation}
F(t)=\langle \hat{B}^\dagger(t)\hat{A}^\dagger(0)\hat{B}(t)\hat{A}(0)\rangle_\beta,
\label{otoc}
\end{equation}
is proposed as a quantum generalization of a classical measure of chaotic behaviors ~\cite{Larkin,Kitaev1}. Here $\hat{H}$ is the system Hamiltonian and $\hat{B}(t)=e^{i\hat{H}t}\hat{B}e^{-i\hat{H}t}$, and $\langle ...\rangle_\beta$ denotes averaging over a thermal ensemble at temperature $1/\beta=k_\text{B}T$.  For a many-body system with local operators $\hat{A}$ and $\hat{B}$, the exponential deviation from unity of a normalized OTOC, i.e. $F(t)\sim 1-\# e^{\lambda_{\text{L}} t}$, gives rise to the Lyapunov exponent $\lambda_\text{L}$.

Quite remarkably, it is found recently that the OTOC also emerges in a different system that seems unrelated to chaos, that is, the scattering of shock waves nearby the horizon of a black hole and the information scrambling there ~\cite{bh1,bh2,bh3}. A Lyapunov exponent of $\lambda_\text{L}=2\pi/\beta$ is found there. Later it is also found that the quantum correction from the string theory always makes the Lyapunov exponent smaller \cite{bh3}. Thus it leads to a conjecture that $2\pi/\beta$ is an upper bound of the Lyaponuv exponent, which is later proved for generic quantum systems \cite{MSS}. This is a profound theoretical result. If a quantum system is exactly holographic dual to a black hole, its Lyapunov exponent will saturate the bound; and a more nontrivial speculation is that if the Lyapunov exponent of a quantum system saturates the bound, it will possess a holographic dual to a gravity model with a black hole. A concrete quantum mechanics model, now known as the Sachdev-Ye-Kitaev model, has been shown to fulfill this conjecture \cite{Kitaev1,Kitaev2,Maldacena1}. This establishes a profound connection between the existence of holographic duality and the chaotic behavior in many-body quantum systems \cite{Shen}.

Recent studies also reveal that the OTOC can be applied to study physical properties beyond chaotic systems. The decay of the OTOC is closely related to the delocalization of information and implies the information-theoretic definition of scrambling. In the high temperature limit (i.e. $\beta=0$), connection between the OTOC and the growth of entanglement entropy in quantum many-body systems has also been discovered quite recently \cite{Zhai,Yoshida}. The OTOC can also characterize many-body localized phases, which are not even thermalized~\cite{MBL1,Zhai,MBL3,MBL4,MBL5}.

Despite of the significance of the OTOC revealed by recent theories, experimental measurement of the OTOC remains challenging. First of all, unlike the normal time-ordered correlators, the OTOC cannot be related to conventional spectroscopy measurements, such as ARPES, neutron scattering, through the linear response theory. Secondly, direct simulation of this correlator requires the backward evolution in time, that is, the ability of completely reverse the Hamiltonian, which is extremely challenging. One experimental approach closely related to time-reversal of  quantum systems is the echo  technique \cite{Hahn50}, and the echo has been studied extensively for both non-interacting particle systems and many-body systems to characterize the stability of quantum evolution in the presence of perturbations \cite{AKD03,QSLZS06, GJPW16} and the physics is already quite close to OTOC. Recently it has been proposed that the OTOC can be measured  using echo techniques \cite{otocexp1}. In addition, there also exists several other theoretical proposals based on the interferometric approaches~ \cite{otocexp2,otocexp3,danshita2016creating}. However, none of them has been experimentally implemented so far.

Here, we adopt a different approach to measure the OTOC. To make our approach work, some extent of  ``local control" is required.  A universal quantum computer fulfills this need with having a ``full local control" of the system -- that is, a universal set of local evolutions can be realized, and this set of local evolutions can build up any unitary evolution of the many-body system, both forward and backward evolution in time. That is to say, we shall use a quantum computer to perform the measurement of the OTOC. In fact, historically, one of the key motivations to develop quantum computers is to simulate the dynamics of many-body quantum systems~\cite{Feynman82}, and quantum simulation of many-body dynamics has been theoretically shown to be efficient with practical algorithms proposed~\cite{Lloyd96}. Here the quantum computer we use is liquid-state NMR with molecules. In this work, we report measurements of OTOCs on a NMR quantum simulator. We should stress that, on one hand, our approach is universal and can be applied to any system that has full local quantum control, including superconducting qubit and trapped-ion; on the other hand, this experiment is currently limited to a small size not because of our scheme but because of the scalability issue of quantum computer.  

\section{NMR Quantum Simulation of the OTOC}

The system we will simulate
is an Ising spin chain model, whose Hamiltonian is written as
\begin{equation}
\hat{H}=\sum\limits_{i}\left(-\hat\sigma^z_{i}\hat\sigma^z_{i+1}+g\hat\sigma^x_{i}+h\hat\sigma^z_{i}\right), \label{Ising}
\end{equation}
where $\hat\sigma^{x,y,z}_i$ are Pauli matrices on the $i$-site. The parameter values $g=1$, $h=0$ correspond to the traverse field Ising model, where the system is integrable. The system is non-integrable whenever both $g$ and $h$ are non-zero. We simulate the dynamics governed by the system Hamiltonian $\hat{H}$, and measure the OTOCs of operators that are initially acting on different local sites. The time dynamics of the OTOCs are observed, from which entanglement entropy of the system and butterfly velocities of the chaotic systems are extracted.

\begin{figure}[b]
\centering
\includegraphics[width=\linewidth]{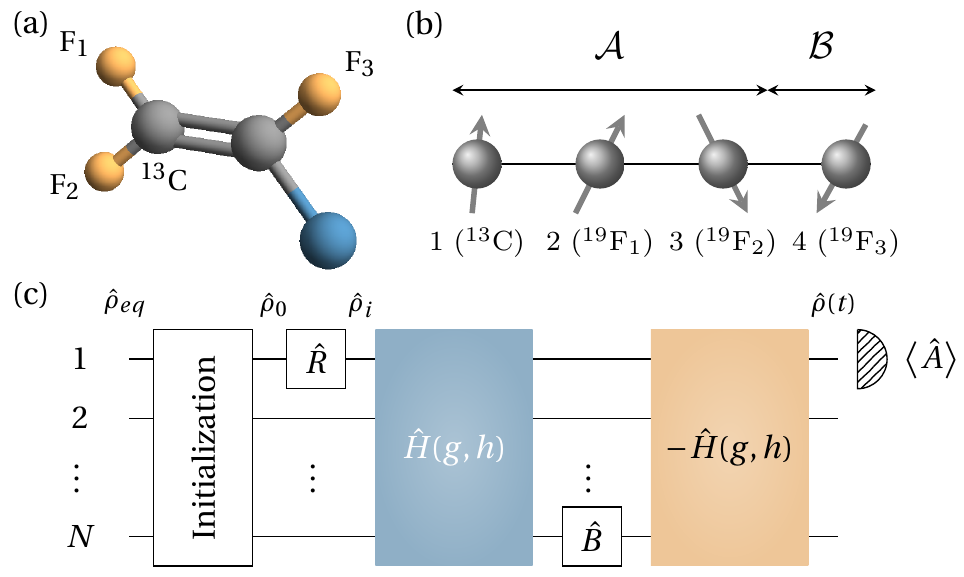}
\caption{Illustration of the physical system, the Ising model and the experimental scheme. (a) The structure of the $\text{C}_2\text{F}_3\text{I}$ molecule used for the NMR simulation. (b) The four sites Ising spin chain, $\mathcal{A}$ and $\mathcal{B}$ label dividing the entire system into two subsystems in the later discussion of entanglement entropy. (c) Quantum circuit for measuring the OTOC for general $N$-site Ising chain when $\beta = 0$ (in our case $N=4$). Here $\hat R = \mathbf{1}, \hat R_x(-\pi/2), \hat R_y(\pi/2)$ for $\hat A = \hat \sigma_1^z, \hat\sigma_1^y, \hat\sigma_1^x$, respectively. }
\label{Schematic}
\end{figure}

\subsection{Physical System}

The physical system to perform the quantum simulation is the ensemble of nuclear spins provided by Iodotrifluroethylene ($\text{C}_2\text{F}_3\text{I}$) which is dissolved in $d$-chloroform, see Fig. \ref{Schematic}(a) for the sample's  molecular structure. For this molecule, the  $^{13}$C nucleus and the three $^{19}$F nuclei ($^{19}$F$_{1}$, $^{19}$F$_{2}$ and $^{19}$F$_{3}$) constitute a four-qubit quantum simulator. Each nucleus corresponds to a spin site of the Ising chain, as shown in Fig. \ref{Schematic}(b). In experiment, the sample is placed in a static magnetic field along $\hat z$ direction,  resulting in the following form of system Hamiltonian
\begin{equation}
\hat{H}_{\text{NMR}}= - \sum_{i=1}^{4}\frac{\omega _{0i}}{2}\hat{\sigma} _{i}^{z}+\sum_{i<j,=1}^{4}%
\frac{\pi J_{ij}}{2}\hat{\sigma} _{i}^{z}\hat{\sigma} _{j}^{z},
\end{equation}
where $\omega_{0i}/2\pi$ is the Larmor frequency of spin $i$, $J_{ij}$ is the coupling strength between spins $i$ and $j$. The values of these system parameters are  given in Appendix \ref{parameters}.
The system is controlled by radio-frequency (r.f.) pulses, and the corresponding control Hamiltonian goes
\begin{equation}
\hat{H}_{\text{rf}} (t)= - \omega_1(t)  \left[ \cos( \phi(t) ) \hat{\sigma} _{i}^{x}  + \sin(\phi (t)) \hat{\sigma} _{i}^{y} \right],
\end{equation}
where $\omega_{1}(t)$ and $\phi(t)$ denote the amplitude and  the emission phase of the r.f. field respectively. The control pulse shape can be elaborately monitored to realize  desired dynamic evolution.
Actually, such a system has been demonstrated complete controllability~\cite{schirmer2001complete}, which guarantes that  arbitrary system evolution can be implemented on it.
Our experiments were carried out on a Bruker AV-400MHz spectrometer ($9.4$ T) at  temperature $T=305$ K.

\begin{figure*}
\centering
\includegraphics[width=0.8\linewidth]{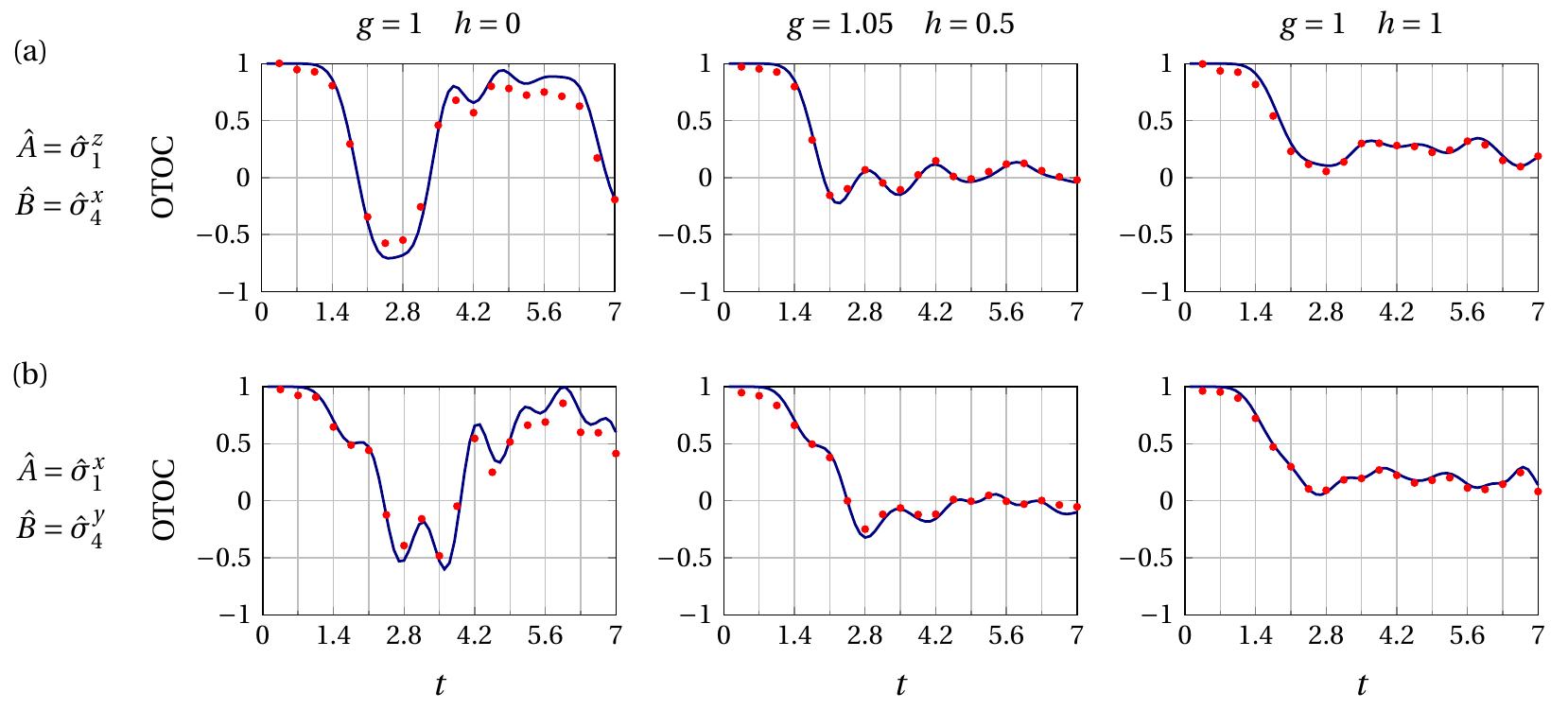}
\caption{Experimental results of OTOC measurement for an Ising spin chain: (a) $\hat{A}=\hat\sigma^z_1$ at the first site, and $\hat{B}=\hat\sigma^x_4$ at the fourth site. (b) $\hat{A}=\hat\sigma^x_1$ at the first site, and $\hat{B}=\hat\sigma^y_4$ at the fourth site. The three columns correspond to $g=1$, $h=0$; $g=1.05$, $h=0.5$; and $g=1$, $h=1$ of model Eq.~\eqref{Ising}, respectively. The red points are experimental data, the blue curves are theoretical calculation of OTOC with model Eq.~\eqref{Ising} for four sites.}
\label{ExpOTOC}
\end{figure*}

\subsection{Experimental Procedure}

As schematically illustrated in Fig. \ref{Schematic}(c), here we focus on $\beta=0$ case and measuring OTOC mainly consists of the following parts.

1. Initial state preparation. This step aims at preparing an initial state with density matrix $\hat \rho_i \propto  \hat{A} =\hat\sigma_1^{\alpha}, \alpha = x, y \text{ or } z$.

1.1. The natural system is originally in the thermal equilibrium state $\hat\rho_{\text{eq}}$ populated according to the Boltzmann distribution. In high-temperature approximation, $\hat\rho_{\text{eq}} \approx 1/2^4(\mathbf{1} + \sum_{i=1}^4 {\epsilon_i \hat\sigma^z_i})$, where $\mathbf{1}$ is the identity and $\epsilon_i \sim 10^{-5}$ denotes the equilibrium polarization of spin $i$. Because there is no observable and unitary dynamical effect on $\mathbf{1}$,  effectively we write $\hat\rho_{\text{eq}} = \sum_{i=1}^4 \epsilon_i{\hat\sigma^z_i}$.

1.2. We engineer the system from $\hat\rho_{\text{eq}}$ into $\hat \rho_0=\sigma^z_1$. This is accomplished in two steps: first to remove the polarizations of the spins except for that of F$_2$ by using selective saturation pulses, and then to transfer the polarization from F$_2$ to $^{13}$C. Details   of the method are described in Appendix \ref{procedure}.

1.3. For initial state $\hat \rho_0$ with $\alpha=x,y$, we need to further rotate spin at site-1 by $\pi/2$ pulse around $y$ or $-x$ axes, respectively.

2. Implementing unitary evolution of $\hat{U}(t)=e^{i\hat{H}t}\hat{B}e^{-i\hat{H}t}$. The key point is that according to the Trotter formula~\cite{Lloyd96}, the time evolution $e^{-i\hat{H}t}$ of the Ising spin chain of Eq.~\eqref{Ising} can be approximately simulated through the decomposition
\begin{equation}
e^{-i \hat{H} m\tau} \approx  \left( e^{-i \hat{H}_x \tau/2} e^{-i \hat{H}_z \tau/2}  e^{-i \hat{H}_{zz} \tau}  e^{-i \hat{H}_z \tau/2} e^{-i \hat{H}_x \tau/2} \right)^m
\label{sim}
\end{equation}
for small enough $\tau$. Here the dynamics is divided into $m$ pieces with $t = m\tau$, and
\begin{subequations}
\begin{align}
&\hat{H}_{x} = \sum\nolimits_{i} g \hat\sigma_i^{x}  \\
&\hat{H}_{z} = \sum\nolimits_{i} h \hat\sigma_i^{z}  \\
&\hat{H}_{zz} = \sum\nolimits_{i} -\hat\sigma^z_{i}\hat\sigma^z_{i+1}.
\end{align}
\end{subequations}
Each propagator inside the bracket of Eq.~\eqref{sim} corresponds to either single-spin operation or coupled two-spin operation, and can be implemented through manipulating $\hat{H}_\text{NMR}$ with r.f. control $\hat{H}_{\text{rf}}$: single-spin operation terms are global rotations around $x$ or $z$ axis, which can be easily done through hard pulses; two-spin operation term $e^{-i\hat{H}_{zz}\tau}$ can be generated through some suitably designed pulse sequence based on the NMR refocusing techniques \cite{Chuang00}.  More details of the method are described in Appendix \ref{procedure}. The reversal of Ising dynamics $e^{i\hat{H}t}$ can be done in the similar manner.  Note in the case considered here, $\hat{B}$ is a local unitary operator on the site-$N$ spin and $\hat{B}=\hat\sigma^\gamma_N$ with $\gamma=x,y,z$ that can be implemented by a selective r.f. $\pi$ pulse on the site-$N$ spin. Hence, for any given $t$, the total unitary evolution $e^{i\hat{H}t}\hat{B}e^{-i\hat{H}t}$ can be simulated.

3. Readout. The OTOC is obtained by measuring the expectation value of the observable $\hat O = e^{i \hat H t} \hat B e^{-i \hat Ht} \hat A e^{i \hat H t} \hat B e^{-i \hat Ht} \hat A$. For the infinite temperature $\beta = 0$, the equilibrium state of the many-body system $\hat{H}$ is the maximally mixed state $\mathbf{1}/2^4$. Since
\begin{equation}
\langle \hat O \rangle_{\beta=0} = \operatorname{Tr} \left(\hat U(t) \hat \rho_0 \hat U^\dag (t) \hat A \right), \end{equation}
when $\hat B$ is unitary, $\hat U(t) \hat  \rho_0 \hat U^\dag (t)$ is a density matrix $\rho(t)$ evolved from $\rho_0$ by $\hat{U}(t)$, as simulated in step 2. Finally $\langle \hat O \rangle_{\beta=0}$ becomes measuring the expectation value of $\hat A$ under $\rho(t)$.
Because that  NMR detection is performed on a bulk ensemble of molecules,    readout is an ensemble--averaged macrosopic measurement. When the system is prepared at state $\rho(t)$, the expectation value of $\hat A$ can then be directly obtained from the  spectrum. See Appendix \ref{procedure} for details.

\subsection{Results of OTOC}

Two sets of typical experimental results of the OTOC at $\beta=0$ are shown in Fig.~\ref{ExpOTOC}. Here we normalize the OTOC by $\langle \hat{B}^\dag(0)\hat{B}(0)\rangle\langle \hat{A}^\dag(0)\hat{A}(0)\rangle$, and because $\hat{A}$ and $\hat{B}^\dag$ commute at $t=0$, the initial value of this normalized OTOC is unity. The experimental data (red points) agree very well with the theoretical results (blue curves). The sources of experimental errors include imperfections in state preparation, control inaccuracy, and decoherence. See Appendix \ref{erroranalysis} for more details. We also measure OTOC for other operators ($\hat{A}=\hat\sigma^\alpha_{1}$, $\hat{B}=\hat\sigma^\gamma_4$ with $\alpha,\gamma=x,y,z$) and they all behave similarly. The experimental results are put in Appendix \ref{procedure}.

In both the integrable case (the first column in Fig.~\ref{ExpOTOC}) and the non-integrable cases (the second and the third columns in Fig.~\ref{ExpOTOC}), the early time behaviors look similar. That is, the OTOC starts to deviate from unity after a certain time (for the unit of time $t$, See Appendix \ref{unit} for details.).  However, the long time behaviors are very different between the integrable and non-integrable cases. In the integrable case, after the decreasing period, the OTOC revives and recovers unity. This reflects that the system has well-defined quasi-particle. And there exists extensive number of integral of motions, which is related to the fact that an integrable system does not thermalize. While in the non-integrable case, the OTOC decreases to a small value and oscillates, which will not revive back to unity in a practical time scale. This relates to the fact that the information does scramble in a non-integrable system~\cite{Yoshida}.

\section{Entropy Dynamics}

To better illustrate the different behaviors of the information dynamics in the two cases of integrable and non-integrable systems, we reconstruct the entanglement entropy of a subsystem from the measured OTOCs. Entanglement entropy has become an important quantity not only for quantum information processing, but also for describing a quantum many-body system, such as quantum phase transition, topological order and thermalization. However, measuring entanglement entropy is always challenging~\cite{guhne2009entanglement}.

OTOC opens a new door for entropy measurement. An equivalence relationship between OTOCs at equilibrium and the growth of the 2nd R\'enyi entropy after a quench has recently been established~\cite{Zhai}, which states that
\begin{eqnarray}
	\exp(-S^{(2)}_\mathcal{A})=\sum_{\hat{M}\in \mathcal{B}}\langle\hat{M}(t)\hat{V}(0)\hat{M}(t)\hat{V}(0)\rangle_{\beta=0}. \label{Theorem}
\end{eqnarray}
In the left-hand side of Eq.~\eqref{Theorem}, $S^{(2)}_\mathcal{A}$ is the 2nd R\'enyi entropy of the subsystem $\mathcal{A}$,
after the system is quenched by an operator $\hat{O}$ at time $t=0$. That is, $S^{(2)}_\mathcal{A}=-\log \hat\rho^2_\mathcal{A}$ and
$
\hat\rho_\mathcal{A}=\text{Tr}_\mathcal{B} (e^{-i\hat{H}t} V e^{i\hat{H}t}),
$ and $\hat{V}=\hat{O}\hat{O}^\dag$, up to a certain normalization condition (see Appendix \ref{normalization}).
The right-hand side of Eq.~\eqref{Theorem} is a summation over OTOCs at equilibrium. $\hat{M}$ is a complete set of operators in the subsystem $\mathcal{B}$.

\begin{figure}[t]
\centering
\includegraphics[width=0.9\linewidth]{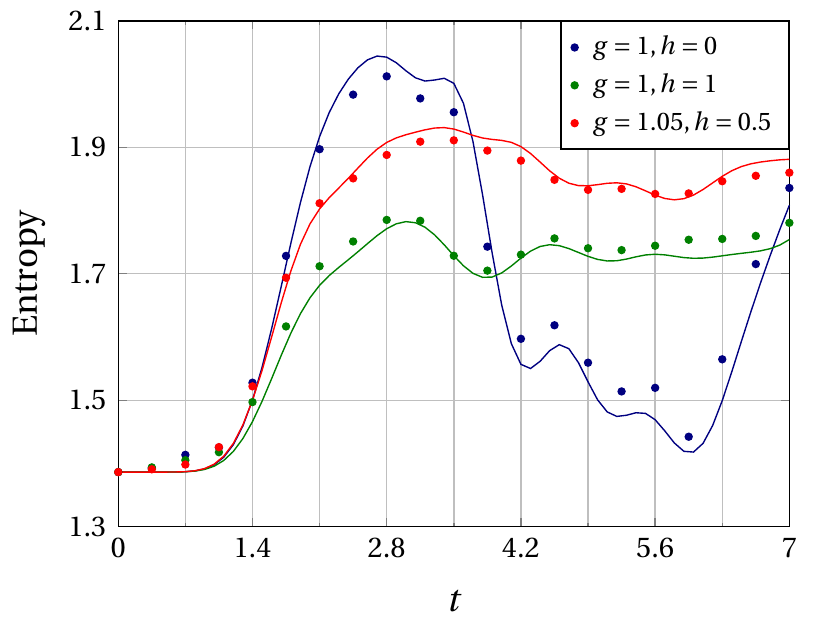}
\caption{The 2nd R\'enyi entropy $S^{(2)}_\mathcal{A}$ after a quench. A quench operator $(\mathbf{1}+\hat\sigma^x_1)$ (up to a normalization factor) is applied to the system at $t=0$, and the entropy is measured by tracing out the fourth site as the subsystem $\mathcal{B}$. Different colors correspond to different parameters of $g$ and $h$ in the Ising spin model. The points are experimental data, the curves are theoretical calculations.}
\label{Entropy}
\end{figure}

In our experiment, we choose the quench operator $\hat{O}\propto(\mathbf{1}+\hat\sigma_1^x)$ at the first site, and we take the first three sites as the subsystem $\mathcal{A}$ and the fourth site as the subsystem $\mathcal{B}$, as marked in Fig.~\ref{Schematic}(b). In this setting, $S^{(2)}_\mathcal{A}$ measures how much the quench operation induces additional correlation between the subsystems $\mathcal{A}$ and $\mathcal{B}$.

We take a complete set of operators in the subsystems $\mathcal{B}$  as $\hat\sigma^\alpha_4$ (up to a normalization factor), where $\alpha=0,x,y,z$ and $\hat\sigma^0=\mathbf{1}$. Since $\hat{V}=\hat{O}\hat{O}^\dag\propto(\mathbf{1} + \hat\sigma_1^x)$, the right-hand side of Eq.~\eqref{Theorem} becomes a set of OTOCs that are given by
\begin{equation}
\label{eq:OTOCq}
\langle\hat\sigma_4^\alpha(t) (\mathbf{1}+\hat\sigma_1^x) \hat \sigma_4^\alpha(t) (\mathbf{1}+\hat\sigma_1^x)\rangle_{\beta=0}.
\end{equation}
Notice that $\text{Tr}(\hat\sigma_4^\alpha(t)\hat\sigma_1^x \hat\sigma_4^\alpha(t))=\text{Tr} (\hat\sigma_4^\alpha(t) \hat\sigma_4^\alpha(t) \hat\sigma_1^x)=0$, the nonzero terms in Eq.~\eqref{eq:OTOCq} are nothing but OTOCs with $\hat{B}=\hat\sigma^\alpha_4$ ($\alpha=x,y,z$) and $\hat{A}=\hat\sigma^x_1$, which are exactly what we have measured. That is to say, with the help of the relationship between OTOCs and entanglement growth, we can extract the growth of the entanglement entropy after the quench from the experimental data.

The results of 2nd R\'enyi entropy $S^{(2)}_\mathcal{A}$ are shown in Fig.~\ref{Entropy}. At short time, all three curves start to grow significantly after certain time. This demonstrates that it takes certain time for the perturbation applied at the first site to propagate to the subsystem $\mathcal{B}$ at the fourth site (see the discussion of butterfly velocity below). Then, for all three cases, $S^{(2)}_\mathcal{A}$s grow roughly linearly in time. This indicates that the extra information caused by the initial quench starts to scramble between subsystems $\mathcal{A}$ and $\mathcal{B}$. The differences lie in the long-time regime. For the integrable model, the $S^{(2)}_\mathcal{A}$ oscillates back to around its initial value after some time, which means that this extra information moves back to the subsystem $\mathcal{A}$ around that time window. As a comparison, such a large amplitude oscillation does not occur for the two non-integrable cases and the $S^{(2)}_\mathcal{A}$s saturate after growing. This supports the physical picture that the local information moves around in the integrable model, while it scrambles in the non-integrable models~\cite{Yoshida}.

\begin{figure}[t]
\centering
\includegraphics[width=0.9\linewidth]{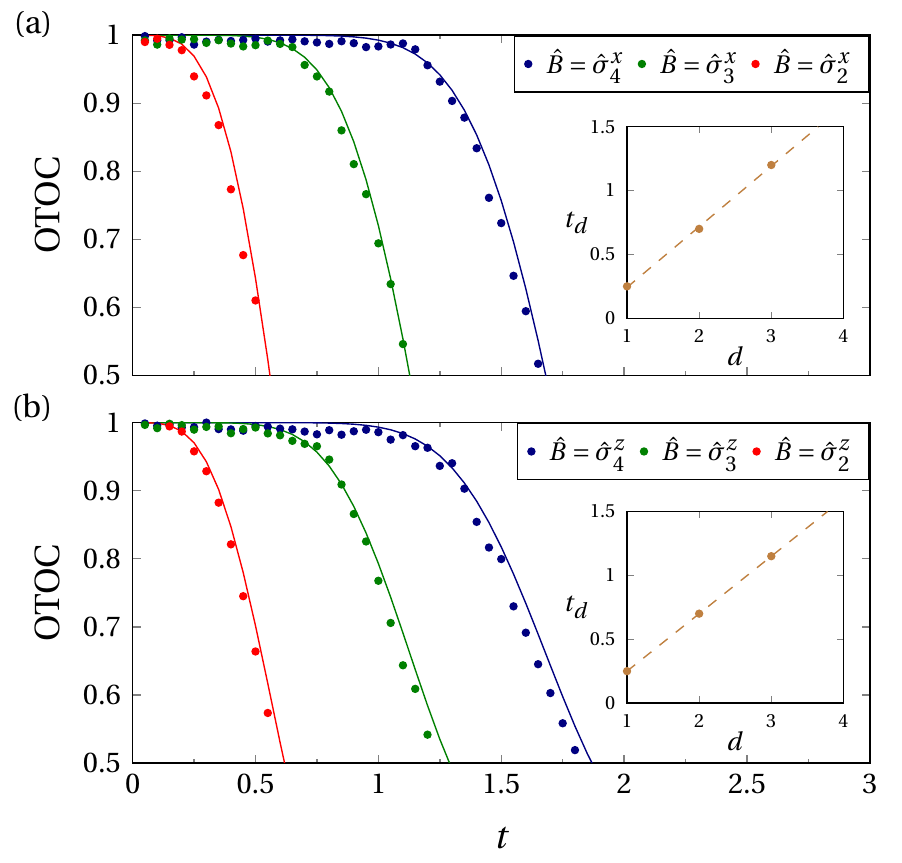}
\caption{Measurement of the butterfly velocity: (a) shows the OTOCs for $\hat{A}=\hat\sigma^z_1$ and $\hat{B}=\hat\sigma^x_i$ with $i=4$ (blue), $i=3$ (green) and $i=2$ (red); (b) shows the OTOCs for $\hat{A}=\hat\sigma^y_1$ and $\hat{B}=\hat\sigma^z_i$ with $i=4$ (blue), $i=3$ (green) and $i=2$ (red). The insets of (a) and (b) shows the time for the onset of chaos $t_d$ for the OTOCs v.s. the distance between two operators. The slope gives $1/v_\text{B}$. Here $g=1.05$ and $h=0.5$. }
\label{BV}
\end{figure}

\section{The Butterfly Velocity}

The OTOC also provides a tool to determine the speed for correlation propagating. At $t=0$, $\hat{A}$ and $\hat{B}$ commute with each other since they are operators at different sites. As time grows, the higher order terms in the Baker-Campbell-Hausdorff formula
\begin{equation}
\hat{B}(t)=\sum\limits_{k=0}^{\infty}\frac{(it)^{k}}{k!}[H,\dots,[H, B],\dots]
\end{equation}
becomes more and more important and some terms fail to commute with $\hat{A}$, at which the normalized OTOC starts to drop. Thus, the larger the distance between sites for $\hat{A}$ and $\hat{B}$, the later time the OTOC starts deviating from unity. In general, the OTOC behaves as
\begin{equation}
F(t)=a-be^{\lambda_\text{L}(t-|x|/v_\text{B})}+\dots,
\end{equation}
where $a$ and $b$ are two non-universal constants, $|x|$ denotes the distance between two operators. Here $v_\text{B}$ defines the butterfly velocity \cite{Yoshida,butterfly,bh3,Blake,Brain}. It quantifies the speed of a local operator growth in time and defines a light cone for chaos, which is also related to the Lieb-Robinson bound \cite{Lieb,Brain}.

In our experiment, we fix $\hat{A}$ at the first site, and move $\hat{B}$ from the fourth site to the third site, and to the second site. From the experimental data, we can phenomenologically determine a characteristic time $t_d$ for the onset of chaos in each OTOC, i.e. the time that the OTOC starts departing from unity. By comparing the three different OTOCs in Fig. \ref{BV}, it is clear that the closer the distance between $\hat{A}$ and $\hat{B}$, the smaller $t_d$. In the insets of Fig. \ref{BV}(a) and (b), we plot $t_d$ as a function of the distance, and extract the butterfly velocity from the slope. We find that, for OTOC with $\hat{A}=\hat\sigma^z_1$ and $\hat{B}=\hat\sigma^x_i$, $v_\text{B}=2.10$; and for OTOC with $\hat{A}=\hat\sigma^y_1$ and $\hat{B}=\hat\sigma^z_i$, $v_\text{B}=2.22$. The butterfly velocity is nearly independent of the choice of local operators, which is kind of manifestation of the chaotic behaviour of the system.

\section{Outlook}

OTOC provides a faithful reflection of the information scrambling and chaotic behaviour of quantum many-body systems. It goes beyond the normal order correlators studied in linear response theory, which only capture the thermalization behaviour of the system. Measuring the OTOC functions can reveal how quantum entanglement and information scrambles across all of the degrees of freedom in a system. In the future it will be possible to simulate more sophisticated systems that may possess  holographic duality, with larger size and different $\beta$, to extract the corresponding Lyapunov exponents such that one can experimentally verify the connection between the upper bound of the Lyapunov exponent and the holographic duality.

We have used liquid-state NMR as a quantum simulator for the demonstration of  OTOC measurement. NMR  provides an excellent platform to benchmark the measurement ideas and techniques.
Our work here represents a first and   encouraging step towards further experimentally observing OTOCs  on large-sized quantum systems.
The present  method can be readily translated to other controllable systems. For instance, in
trapped-ion systems there have been realized high-fidelity execution of arbitrary control with up to five atomic ions  \cite{Ion16-1}. Superconducting quantum circuits also allow for engineering on local qubits with errors at or below the threshold \cite{Barends14,Kelly15}, hence offering another very promising experimental approach. 
Recent years'  progress in these   two quantum hardware platforms has been fast and astounding, particularly in the pursuit of fabrication of  quantum  computing architecture  at large scale. It is reckoned   that    quantum simulators   consisting of tens of or even hundreds of qubits  are within reach  in the near future \cite{MK13,Ion16-2,Chow15,GCS17}. Experimentalists will see the great opportunity of applying these technologies for studying quantum chaotic behaviors     for  much more complicated quantum many-body systems.

\section{Acknowledgments}

We thank Huitao Shen, Pengfei Zhang, Yingfei Gu and Xie Chen for helpful discussions. B. Z. is supported by NSERC and CIFAR. H. Z. is supported by MOST (grant no. 2016YFA0301604), Tsinghua University Initiative Scientific Research Program, and NSFC Grant No. 11325418.   H. W., X. P., and J. D. would like to thank the following funding sources: NKBRP (2013CB921800 and 2014CB848700), the National Science Fund for Distinguished Young Scholars (11425523), NSFC (11375167, 11227901 and 91021005). J. L. is   supported by the National Basic Research Program of China (Grants No. 2014CB921403, No. 2016YFA0301201), National Natural Science Foundation of China (Grants No. 11421063, No. 11534002, No. 11375167 and No. 11605005), the National Science Fund for Distinguished Young Scholars (Grant No. 11425523), and NSAF (Grant No. U1530401).

\emph{Note added.}---
After finishing this work, we notice a related work~\cite{Martin16}, where OTOCs are measured in a
trapped ion quantum magnet.

\appendix

\section{Parameters of the System Hamiltonian}
\label{parameters}

\begin{figure}[t]
\centering
\includegraphics[width=\linewidth]{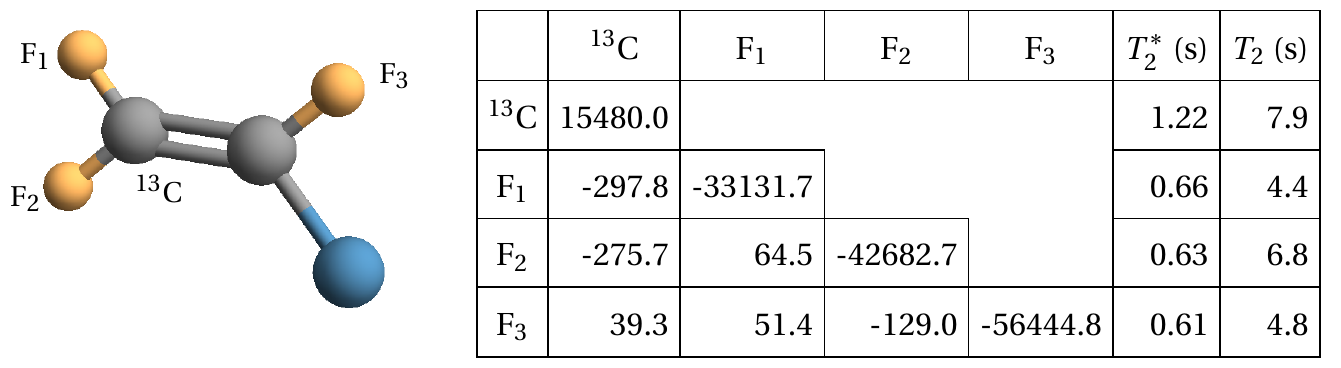}
\caption{Characteristics of Iodotrifluroethylene. Molecular structure  together with a table of the chemical shifts (on the diagonal) and $J$-coupling strengths (lower off-diagonal), all in Hz. The chemical shifts are given with respect to base frequency for $^{13}$C or $^{19}$F  transmitters on the  400 MHz spectrometer that we used. }
\label{CFFF}
\end{figure}

We use Iodotrifluroethylene dissolved in d-chloroform~\cite{peng2014experimental}. The system Hamiltonian is given by
\begin{equation}
\hat{H}_{\text{NMR}}= - \sum_{i=1}^{4}\frac{\omega _{0i}}{2}\hat{\sigma} _{i}^{z}+\sum_{i<j,=1}^{4}%
\frac{\pi J_{ij}}{2}\hat{\sigma} _{i}^{z}\hat{\sigma} _{j}^{z},
\end{equation}
where $\omega_{0i}/2\pi$ is the Larmor frequency of spin $i$, $J_{ij}$ are the coupling strength between spins $i$ and $j$. The values
of parameters $\omega_{0i}$ and $J_{ij}$ are given in Fig.~\ref{CFFF}.

\begin{figure*}
\centering
\includegraphics[width=0.95\textwidth]{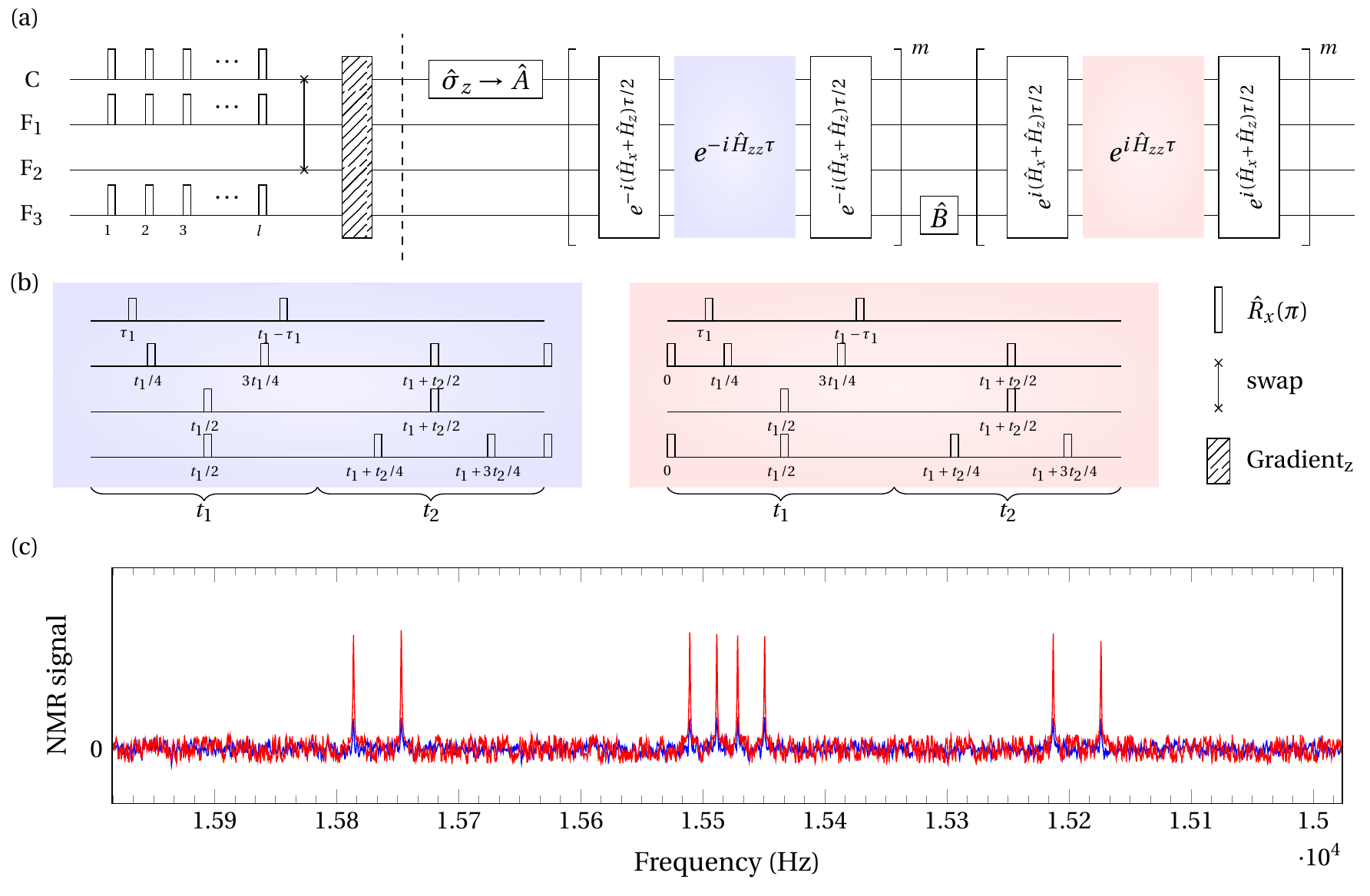}
\caption{
(a) Quantum circuit that  measures the OTOCs. The first part aims at reset an arbitrary state to the desired initial state. Here the time interval between the $\pi$ pulses is $25$ ms, the number of cycles is $l=500$, and G$_z$ denotes $z$ axis gradient pulse.   (b) Sequences for implementing the dynamics of $e^{-i \hat H_{zz}\tau}$ (left) and $e^{i \hat H_{zz}\tau}$ (right). The  refocusing circuits are designed to  generate the right amount of  coupled evolution. (c) $^{13}$C experimental spectrum for equilibrium state (blue) and  state $\hat \rho_0$ (red) after a readout pulse $\hat R^1_y(\pi/2)$.  They are shown at the same scale for comparison.}
\label{Circuit}
\end{figure*}

\section{Experimental Procedure}
\label{procedure}

\subsection{Initialization}
The system is required to be initialized into $\hat \rho_{0}  \propto \hat \sigma_1^z$ from the equilibrium state $\hat \rho_{eq}$. We first exploit the steady state effect when a relaxing nuclear spin system is subjected to  multiple-pulse irradiation~\cite{levitt1992steady}. To implement this, we apply the  periodic sequence  $[\pi_{1,2,4} - d]$ to the system, where $\pi_{1,2,4}$ means simultaneous $\pi$ rotations on the spins $^{13}$C, F$_1$ and F$_3$, and $d$ is a time delay parameter to be adjusted, see the first part of the circuit shown in Fig. \ref{Circuit}(a). To do $\pi_{1,2,4}$, we use a pulse which is composed of three frequency components,  each Hermite-180 shaped in 500 segments, with a duration of 1 ms. With increasing the number of applied cycles, under the joint effects of relaxation and $\pi$ reversions, the equilibrium Zeeman magnetizations $\left\langle \hat \sigma^z_{1,2,4} \right\rangle$ gradually decay to zero.   Only the magnetization $\hat \sigma_3^z$ is retained at last as it  is the fixed point to the periodic driving. We   adjust the time interval $d$ between the $\pi$ pulses to achieve the best-quality steady state. In experiment, we set $d=25$ ms and after more than 500 cycles we found that the system was effectively steered into a steady state $\hat\rho_{ss}  \propto \hat \sigma_3^z$ (in this sample, we did not see observable Overhauser enhancement).
Next, with a SWAP operation we transfer the polarization from the high-sensitivity F$_2$ nucleus to the low-sensitivity $^{13}$C nucleus. Using the method, we finally get an initial state $\hat \rho_{0} \propto  \hat\sigma^z_1$. The resulting experimental spectrum  is shown in Fig. \ref{Circuit}(c).


\subsection{Simulating time evolution of Ising spin chain}
According to Eq. (5) of the main text, the key ingredient in simulating the evolution of Ising Hamiltonian $\hat H$ is to implement
\begin{equation}
e^{-i \hat{H}_x \tau/2} e^{-i \hat{H}_z \tau/2}  e^{-i \hat{H}_{zz} \tau}  e^{-i \hat{H}_z \tau/2} e^{-i \hat{H}_x \tau/2}.
\end{equation}
Here, except for $e^{-i\hat{H}_{zz}\tau}$, all other four terms are global rotation around $x$ (and $z$) axis, which can be easily done through hard pulses. $e^{-i\hat{H}_{zz}\tau}$ can be generated by manipulating the natural physical Hamiltonian $\hat H_{\text{NMR}}$ with a suitable refocusing scheme ~\cite{vandersypen2005nmr}. The basic idea is to evolve the system with the $J$-term in $\hat{H}_\text{NMR}$ and then to use spin echoes  to engineer the evolution. That is to say, for instance, for the $\hat{\sigma}^z_i\hat{\sigma}^z_j$ term, when a transverse $\pi$ pulses is applied to reverse the polarization of one of the two spins, the evolution is also reserved.
Hence by designing a suitable refocusing scheme, the dynamics of $\hat H_{zz}$ and $-\hat{H}_{zz}$  can be efficiently simulated.

Although general and efficient refocusing scheme exists for any $\hat \sigma_z \hat \sigma_z$-coupled evolution~\cite{Chuang00}, for the present task it is possible to find a much simplified circuit construction.
Fig. \ref{Circuit}(b) shows our ideal circuits.
Let $\text{O}_1$ and $\text{O}_2$ define the reference frequency for carbon and fluorine channel respectively. Consider the refocusing circuit (Fig. \ref{Circuit}(b), left) for implementing $e^{-i\hat H_{zz} \tau}$, it automatically refocuses the fluorine spins and decouples the terms $J_{31}$, $J_{41}$ and $J_{43}$, and  the evolution of other  terms should fulfil the following requirements to yield the right amount of evolution:
\begin{subequations}
\begin{align}
 (\omega_{01} /2\pi - \text{O}_1)(4\tau_1 - t_1 +t_2)  & =  0, \\
 \pi J_{21}/2 \times 4\tau_1  & = - \tau, \\
 - \pi J_{32}/2 \times t_2  & = - \tau, \\
  \pi J_{43}/2 \times t_1  & = - \tau.
\end{align}
\end{subequations}
The solution to the above system of equations is given by $\text{O}_1 = \omega_{01}/ 2\pi = 15480.0$ Hz, $t_1 = 0.004935 \tau$, $t_2 = 0.009870 \tau$ and $\tau_1 = 0.000534 \tau$. As to the refocusing circuit for implementing $e^{i\hat H_{zz} \tau}$, we found that it suffices to just make slight changes to the circuit for $- e^{i \hat H_{zz} \tau}$, as shown in the figure, and one can then reverse the dynamics of all terms.

\begin{figure*}
\centering
\includegraphics[width=0.75\textwidth]{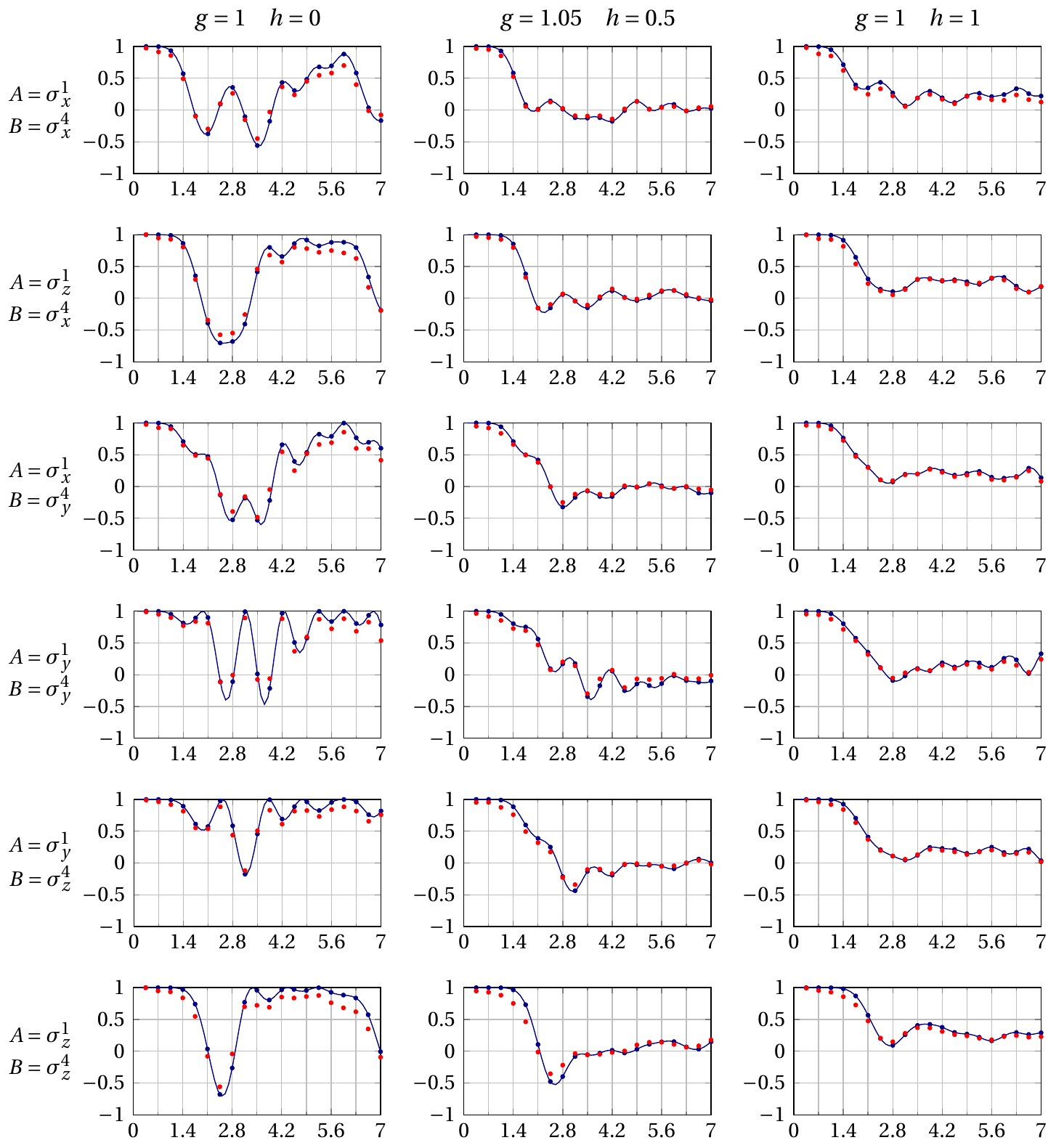}
\caption{Experimental results for measuring OTOCs for different Ising model parameters and different pairs of $\hat A$ and $\hat B$. The red points are experimental data, the blue curves are theoretical calculation of OTOC with model, the blue points are theoretical values displayed for comparison.}
\label{Results}
\end{figure*}

Now, the whole network for implementing Ising dynamics is expressed in terms of single-spin rotations and evolution of $J$-terms in $\hat{H}_\text{NMR}$.
In practices, each single spin rotation is realized through a selective r.f. pulse of Gaussian shape, with a duration of 0.5 to 1 ms.  We then conduct a compilation procedure to the sequence of selective pulses to eliminate the control imperfections caused by off-resonance and coupling effects up to the first order
\cite{Ryan08, jun16}. To further improve the control performance, we employ the gradient ascent pulse engineering (GRAPE) technique \cite{KRKSG05} on the complied sequences. Because that compilation procedure has the capability of directly providing a good initial start for subsequent gradient iteration, the GRAPE searching quickly finds out  high performance pulse controls for the desired propagators.
The obtained shaped pulses for different set of Hamiltonian parameters $(g,h)$ all have  the numerical fidelities above 0.999, and have been optimized with practical control field inhomogeneity taken into consideration.

The Ising dynamics to be simulated is discretized into $20$ steps, with each time step of duration $\tau = 0.35$ ms.
Choosing different operators for $\hat A$ and $\hat B$, we have experimentally measured the corresponding OTOC. All the experimental results are given in Fig.~\ref{Results}. The theoretical  trajectories  are  plotted for comparison. Although some discrepancies between the data and the simulations remain, the experimental results  reflect very well how OTOCs behave differently in the integrable and chaotic cases.

\begin{figure*}
\centering
\includegraphics[width=0.75\textwidth]{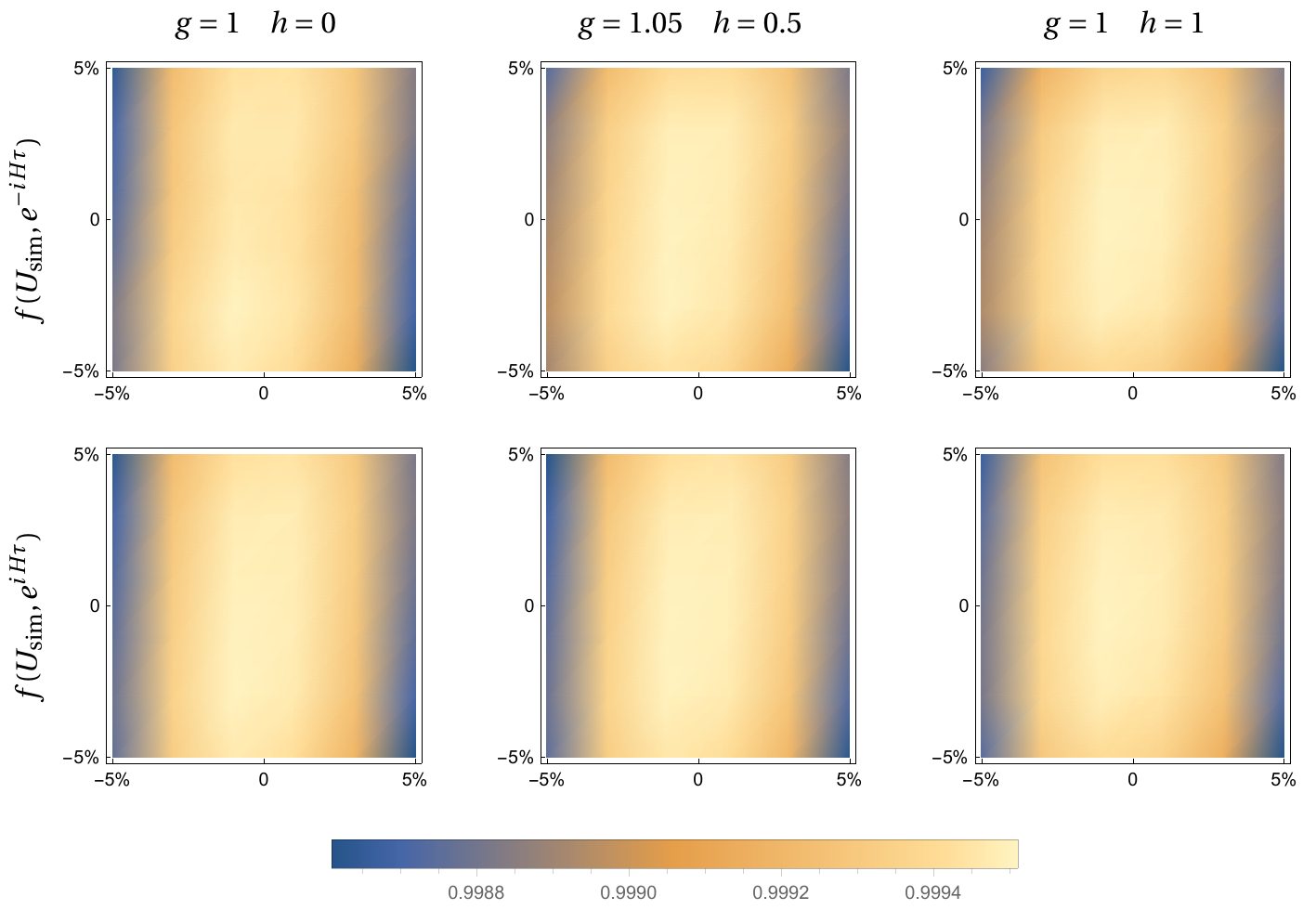}
\caption{Robustness of the used GRAPE pulses against r.f. field inhomogeneity. Here the transverse axis denotes relative error of output field power of $^{13}$C channel, and longitudinal axis that of $^{19}$F channel, $U_\text{sim}$ denotes the corrsponding propagator, $f$ is fidelity function.}
\label{robustness}
\end{figure*}

\subsection{Readout}
All the observations are made on the probe spin $^{13}$C. Because we use an unlabeled sample in real experiment, the molecules with a
$^{13}$C nucleus are present at a concentration of about $1\%$. The NMR signal in high field is obtained from the precessing transverse magnetization of the ensemble of molecules in the sample:
\begin{eqnarray}
M(t) &=& M_x(t) + i M_y(t) \nonumber\\
&=& \text{Tr} \left[ \hat \rho(t) \left(\sum_j  \left\langle \sigma_j^x \right\rangle + i \sum_j  \left\langle \sigma^y_j \right\rangle \right)\right]
\end{eqnarray}
As the precession frequencies of different spins are distinguishable, they can be individually detected, e.g., we obtained the measurements of $\langle \hat \sigma_1^x\rangle$ and $\langle \hat \sigma_1^y\rangle$ at the $^{13}$C Larmor frequency. To measure $\langle \hat \sigma_1^z\rangle$, we need to apply a $\pi/2$ rotation along $\hat{y}$. By fitting the  $^{13}$C  spectrum, the real part and imaginary parts of the peaks are extracted, which corresponds to $\langle \hat \sigma_1^x\rangle$ and $\langle \hat \sigma_1^y\rangle$, respectively.

\section{Experimental Error Analysis}
\label{erroranalysis}

The sources of experimental errors include imperfections in initial state preparation,  infidelities of the GRAPE pulses, r.f. inhomogeneity, and decoherences. We make analysis to the data set of the case $\hat A = \hat \sigma_1^x, \hat B = \hat \sigma_4^y$ to get an understanding on the role of each type of error sources. We calculated the standard deviations $\sigma_\text{exp} := 􏰁\sqrt{ \sum\nolimits_{i=1}^{20} ( \langle\hat A  \rangle^i_\text{exp} - \langle \hat A \rangle^i_\text{th} )^2/20}$ for the experimental data, which are presented in Table \ref{error}.

\begin{table}[b]
\begin{center}
{\small
\renewcommand{\arraystretch}{1.25}\setlength{\tabcolsep}{5pt}
\begin{tabular}{|c|c|c|c|}
\hline
 & $g=1, h=1$ &  $g=1.05,h=0.5$ &  $g=1, h=0$ \\ \hline
$\sigma_\text{exp}$ &  0.1097 &  0.0456  &        0.0308     \\ \hline
$\sigma^\text{err}_\text{ini}$ &  0.0340 &  0.0340 &  0.0340 \\ \hline
$\sigma^\text{err}_\text{inhomo}$ &  0.0323 &  0.0150 &  0.0188 \\ \hline
$\sigma^\text{err}_{T_2}$ &  0.0461 &  0.0161 &  0.0214  \\ \hline
\end{tabular}
}
\end{center}
\caption{The standard deviations of $ \langle \hat A  \rangle$ for the experiments and numerical simulations when $\hat A = \hat \sigma_1^x, \hat B = \hat \sigma_4^y$.}
\label{error}
\end{table}

We have run the initialization process for 50 times and found that the fluctuation of the initial state polarization of $\hat \rho_0$ is around $3.40\%$. The fluctuation is due to (i) error in state preparation; (ii) error in spectrum fitting.  The latter can be inferred from the signal-to-noise ratio of the spectrum, which is estimated to be $\approx 2.13\%$.

All the GRAPE pulses for  implementing $e^{-i \hat H \tau}$ and $e^{i \hat H \tau}$, are of fidelities above 0.999.
On such precision level, if we assume no other sources of error and assume that the pulse generator ideally generates these pulses,  then the experimental results should  match the theoretical predictions almost perfectly.

Fig. \ref{robustness} plots the   robustness of the GRAPE pulses in the presence of imperfections of r.f. fields in the $^{13}$C channel and $^{19}$F channel.
To understand  to what extent the r.f. field inhomogeneity may affect the experimental results, we calculate the  deviation   of the dynamics based on a simple inhomogeneity model. The model assumes that the output power discrepancy  of the r.f. fields is uniformly distributed between $\pm 3\%$. The simulated results $\sigma^\text{err}_\text{inhomo}$ are shown in Table \ref{error}.

Another major source of error comes from  decoherence effects. We compare the experimental data to a
simple phenomenological error model, i.e., the system undergoes uncorrelated dephasing channel, parameterized with a set of phase flip error probabilities $\left\{ p_i \right\}_{i=1,2,3,4}$  per evolution time step $t_0$. The density matrix $\hat \rho$ is then, at each evolution step,
subjected to the composition of the error channels $\mathcal{E}_i$ for each qubit~\cite{NC00}
\begin{equation}
\hat \rho \to \mathcal{E}_4 \circ \mathcal{E}_3 \circ \mathcal{E}_2 \circ \mathcal{E}_1 (\hat\rho),
\end{equation}
where
\begin{equation}
 \mathcal{E}_i (\hat\rho) = (1-p_i) \hat\rho + p_i \hat\sigma^z_i \hat \rho \hat\sigma^z_i.
\end{equation}
with   $p_i = (1 - e^{- t_0/T_{2,i}})/2$ (see Fig. \ref{CFFF} for the values of   $T_{2,i}$). The results are presented in Table \ref{error}.
The results indicate that, with decoherence effects taken into account, the discrepancy between theoretical and experimental data for $g=1, h=0$ is expected to be larger than that of the other two cases, consistent with the experiment data.

In summary, we conclude that r.f. inhomogeneity and decoherence effects are two major sources of errors.

\section{The Unit of Time $t$}
\label{unit}
Our model Hamiltonian is actually written as $\hat H = \sum_i (-J \hat \sigma_i^z \hat \sigma_{i+1}^z + g\hat \sigma_i^x + h\hat \sigma_i^z)$, where we automatically set $J=1$ in the main text. And we choose the natural unit $\hbar = 1$ throughout. So our time $t$ is in fact in the unit of $\hbar/J$.

\section{Normalization Condition for the Entanglement Entropy and OTOC Relation}
\label{normalization}

The   relationship between the growth of 2nd R\'enyi entropy after a quench and the OTOCs at equilibrium is given in~\cite{Zhai}. For a system at infinite temperature, we quench it with any operation $\hat O$ at $t=0$. So the density matrix at time $t$ is $\hat \rho(t) = e^{-i\hat Ht}\hat O\hat {\mathbf{1}}\hat O^\dag e^{i\hat Ht}$ . Then we study the second entanglement R\'enyi entropy between the subregion $\mathcal{B}$ and the rest is denoted as $\mathcal{A}$. The reduced density matrix is $\hat \rho_\mathcal{A}(t) = \text{Tr}_\mathcal{B} \hat \rho(t)$, which gives us the entropy $S_\mathcal{A}^{(2)}(t) = -\log \text{Tr}_\mathcal{A}[\hat \rho_\mathcal{B}(t)^2] $. The growth of entanglement is related to the OTOCs via
\begin{eqnarray}
	\exp(-S^{(2)}_\mathcal{A})=\sum_{\hat{M}\in \mathcal{B}}\langle\hat{M}(t)\hat{V}(0)\hat{M}(t)\hat{V}(0)\rangle_{\beta=0},\label{supplement}
\end{eqnarray}
where the summation is taken over a complete set of operators in $\mathcal{B}$ and $\hat V=\hat O \hat O^\dag$. Here we should choose the following normalization condition: $\sum_{\hat{M}\in B}M_{ij}M_{lm}=\delta_{im}\delta_{lj}$, $\text{Tr}[\hat{O}\hat{O}^\dagger]=\mathbf{\hat 1}$.

\begin{figure}[b]
	\includegraphics[width=0.75\linewidth]{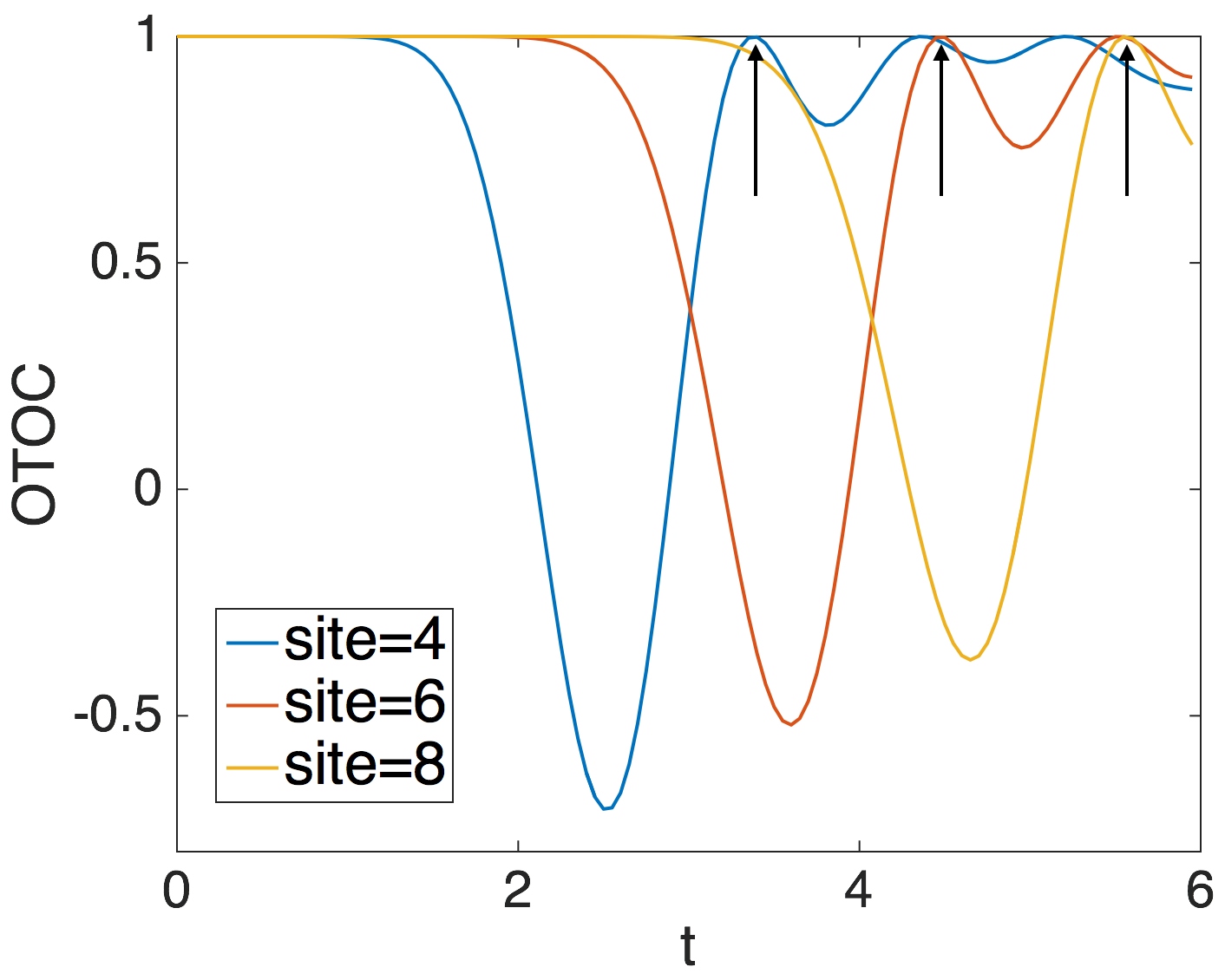}
	\caption{Numerical results of OTOCs for the integrable case. The arrows denote the revival time, which approximately linearly increases with respect to the distance between operators. Here we choose $\hat A = \hat\sigma_z^1$ on the first site and $\hat B = \hat\sigma_z^n$ on the final site. The parameters are $g=1, h=0$.}
	\label{OTOC_revival}
\end{figure}

Here, we quench the first site and take the first three sites as the subsystem $\mathcal{A}$ and the fourth site as the subsystem $\mathcal{B}$, as marked in Fig. 1(b) of the main text. Hence, we choose $\hat{O} = (\mathbf{\hat 1}+\hat \sigma_1^x)/2^{(D+1)/2}$ ($D=4$ is the total number of sites). The complete set of operators in the subsystems $\mathcal{B}$ can be taken as $\hat \sigma^\alpha_4/\sqrt{2}$, where $\alpha=0,x,y,z$ and $\hat \sigma^0=\mathbf{\hat 1}$. By summing over the measured data with the conventions above, we can get the points in Fig. 3 of the main text. The theoretical curves are obtained by directly computing entanglement entropy from the density matrix.

\section{Revival Time of OTOC and the Distance Between the Operators}

As seen from Fig. 2 of the main text, for the integrable case, the OTOCs will increase back around their initial values at some time. The revival time in fact depends on the spatial distance between the two operators, as depicted in Fig.~\ref{OTOC_revival}. That is, the larger the distance, the later the revival happens. From the relationship between  the growth of 2nd R\'enyi entropy after a quench and the OTOCs at equilibrium given in~\cite{Zhai}, we know that it will take longer time for the entanglement entropy to decrease back after a local quench.

\end{document}